# Beam tail effect of performance-enhanced EC-ITC RF gun

HU Tong-ning(胡桐宁)[1], PEI Yuan-ji(裴元吉)[2], QIN Bin(秦斌)[1;1)], CHEN Qu-shan(陈曲珊)[1]

[1] *State Key Laboratory of Advanced Electromagnetic Engineering and Technology, Huazhong University of Science and Technology, Wuhan 430074, China;*
[2] *National Synchrotron Radiation Laboratory, University of Science and Technology of China, Hefei 230029, China*

**Abstract:** Beam tail effect of multi-bunches will influence the electron beam performances in high intensity thermionic RF gun. Beam dynamic calculations that illustrate the working states of single and multi-pulse fed-in of performance-enhanced EC-ITC (External Cathode Independent Tunable Cavity) RF gun for FEL (Free Electron Laser) injector are performed to estimate extracted bunch properties. By using both Parmela and homemade MATLAB codes, the effects of single beam tail as well as interactions of multi-pulses are analyzed, where ring-based electron algorithm is adopted to calculated RF fields and space charge field. Furthermore, the procedure of unexpected deviated-energy particles mixed with effective bunch head is described by MATLAB code as well. As a result, performance-enhanced EC-ITC RF gun is proved to have the capability to extract continual stable bunches which are suitable for high requirement THz-FEL.

**Keyword:** EC-ITC, thermionic RF gun, beam tail effect, multi-pulses, FEL injector, beam dynamics

**PACS:** 29.25.Bx, 41.85.Ar, 29.20.Ej

## 1 Introduction

Nowadays, high quality electron beam sources have been focused due to rapid development of FEL facilities, which have many advantages such as wave-length designable, continuous speedy tunable, high power and high efficiency potentials [1]. Though photocathode RF gun has been widely used as pre-injectors of Linacs for FELs, thermionic RF gun still has the potential of generating high brightness electron beams with high intensity and low emittance suitable for long wavelength FELs [2,3].

The photocathode gun can provide brilliant bunches owing to its high gradient fields, extreme low thermal emittance, as well as the controllable time-structure provided by embed laser pulses. However, this type of electron gun is much more expensive and complex. The thermionic RF gun still plays an important role and is expected to generate high quality bunches, due to its simpler structure and lower cost. It is necessary to improve the performances of thermionic RF gun in order to meet strict requirements of FEL. Researches focused on multi-cavity structures have produced a remarkable effect on the aspect of back-bombard (BB) power inhibition by shortening the first cavity and adjusting electric field strength ratio [4,5,6]. However, the sacrifice of space dimensions is inevitable, which is resulted from using $\alpha$-magnets to compress pulse length. For the sake of compactness, the ITC RF gun without $\alpha$-magnets was proposed in Japan several years ago[7], which has multi-cell structure with independent power fed-in. After that, China Academy of Engineering Physics (CAEP) has got good results by using ITC concept [8]. Though bunches with excellent characteristics such as ~0.2ps pulse length (FWHM) with tens of pC bunch charge could be obtained, the bottleneck of this type of RF gun is that the bunch charge is hard to exceed to over 100pC , due to the back stream from the first cell to embed cathode. To solve this issue, an advanced structure with external injection, so called EC-ITC RF gun is designed and manufactured in National Synchrotron Radiation Laboratory (NSRL)[9,10], which has canceled BB effect almost completely, and enhanced bunch charge to ~130pC with 0.5% energy spreads.

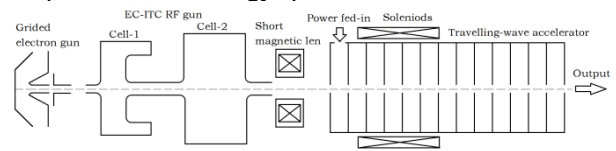

Fig. 1. Schematic of FEL injector.

The schematic of beam injector used for a THz-FEL oscillator proposed in Huazhong University of Science and Technology (HUST)[11,12] is shown as Fig. 1, which mainly consists of grided electron gun, a EC-ITC RF gun, a traveling-wave accelerator and focusing magnets. For physics requirements of the THz-FEL, ~200pC bunch charge with 0.3% energy spread and 5ps bunch length (FWHM) is preferred. Consequently, the input beam for EC-ITC RF gun must be enhanced to ~ 5A DC beam due to the limitation of

1).Corresponding author (email: bin.qin@mail.hust.edu.cn).

capture rate of thermionic RF gun itself. A design modification of the original EC-ITC RF gun was proposed here to enhance the properties of extracted bunches, by using grided electron gun with double anodes as external injection instead of diode gun. Restricted by the bunching capacity in this thermionic RF gun, considerable amount of electrons located at beam tail exist during the whole RF period, which might influence the subsequent pulse properties. Besides, a part of energy-deviated particles mixed in bunch head would be extracted together with effective bunch head. Both influences of these phenomena should be taken into account and analyzed in advance to guarantee the continual stability of extracted effective bunches for FEL injector.

## 2　Structure design of EC-ITC RF gun

As a critical component of FEL injector illustrated in Fig. 1, EC-ITC RF gun has a double-cell structure with independent power fed-in, and based on researches of the multi-cavity RF gun by Kim [13], there're two main criterions for structure design:

1) The length of the drift tube ($L$) between two cells must be large enough to attenuate the electric field strength ($E_L$) to 0.1% of peak value ($E_0$), which could be given by Eq. (1) and Eq. (2) as following,

$$\frac{E_L}{E_0} = e^{-\alpha L}. \quad (1)$$

$$\alpha = \frac{2\pi}{\lambda}\sqrt{\left(\frac{2.405 \cdot \lambda}{2\pi b}\right)^2 - 1}. \quad (2)$$

where, $\alpha$ is the attenuation constant, $\lambda$ is the RF wavelength, and $b$ is the radius of drift tube.

2) The length of standing-wave cavity ($L_i$) should be set appropriately non-relativistic cases, since the relative velocity of the synchronous electron $\beta_{ei} < 1$.

$$L_i = \beta_{pi}\frac{\lambda}{2}. \quad (3)$$

where, $\beta_{pi}$ is the phase velocity of the standing-wave cavity, and must obey $\beta_{ei} \equiv \beta_{pi}$ to guarantee synchronous electrons meet the same field phase when they enter into the cavity.

By adjusting cavity dimensions using Superfish[14], the electric field distributions of both Cell-1 and Cell-2 are shown in Fig. 2, and the electric field strengths on beam axis are plotted by Parmela[15] as well. Apparently, both of the two cells work at $TM_{010}$ mode with 2856MHz resonance frequency, and there's only longitudinal component of electric field strength near the beam axis, thus, beam from grided electron gun can be bunched by Cell-1, accelerated and bunched by Cell-2, then injected into the travelling wave accelerator.

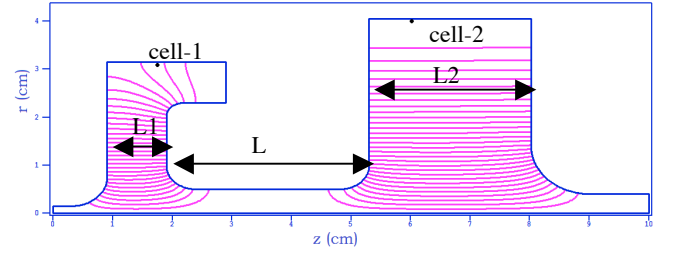

(a) Structure and electric field distributions

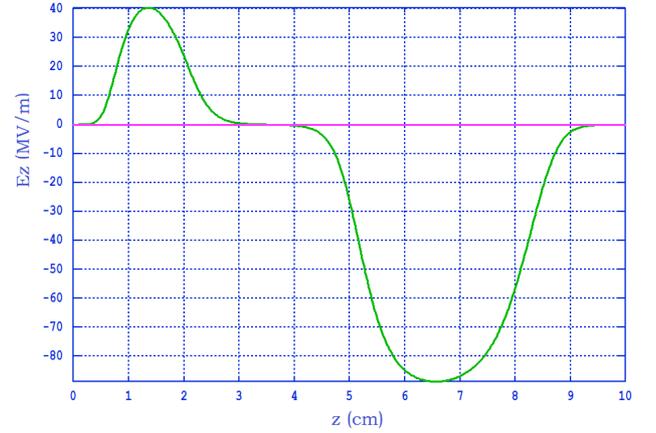

(b) Longitudinal electric field strengths on beam axis.
Fig.2. Calculated electric field lines and longitudinal electric field strengths on beam axis.

## 3　Beam dynamics of single pulse

Since 5A output macro-pulse with 1mm waist radius and 42mm range could be obtained from grided electron gun by adopting double anodes designed by Superfish, its output beam could be used as input file for ITC RF gun directly, and beam envelope calculated Parmela with space charge force considered is shown in Fig. 3; detailed specifications are listed in Table 1 correspondingly.

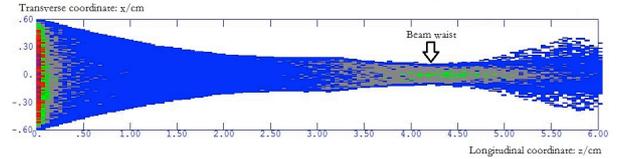

Fig. 3. Beam envelope extracted from grided electron gun.

Table 1　Specifications of extracted beam.

| Parameter | Value |
|---|---|
| Beam kinetic energy | 15keV |
| Beam Intensity | 5A |
| Radius of beam waist | 1mm |

| Transverse normalized emittance (rms) | $5\pi$ mm·mrad |
|---|---|
| Energy spread (rms) | 0.1% |

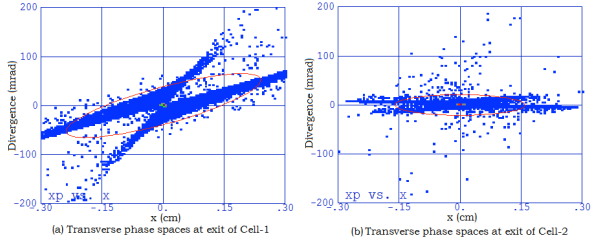

Fig. 4. Transverse phase spaces of EC-ITC-RF gun.

By adjusting feed-in RF parameters such as powers and phases of two independent cells in the EC-ITC RF gun using Parmela, desired beam specifications can be achieved. Fig. 4 shows the both transverse phase spaces located at Cell-1 and Cell-2, while the final working states are illustrated clearly by Fig. 5, and generated bunch properties are listed in Table 2, which indicates short-length and high-brightness bunches have been obtained. The essential parameters of 196pC effective electric charge, 1.3ps bunch length (FWHM) with 0.28% energy spread (FWHM) and 6.5π mm·mrad normalized emittance (rms) of effective bunch head provide good enough beam quality for the FEL injector.

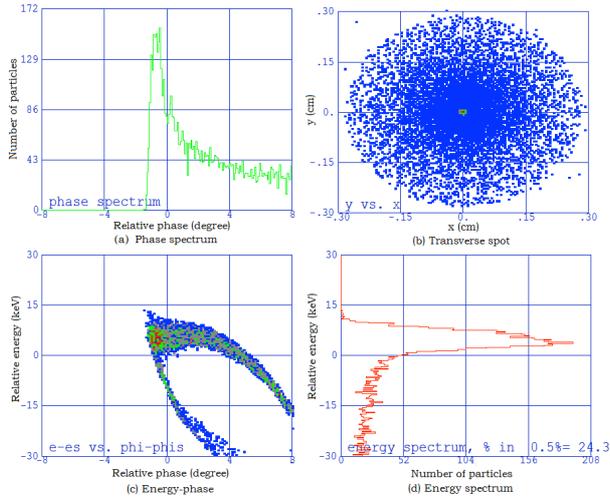

Fig. 5. Output bunch states of EC-ITC-RF gun.

Table 2  Output bunch properties of ITC-RF gun.

| Parameter | Value |
|---|---|
| RF frequency | 2856MHz |
| Beam kinetic energy | 2.59MeV |
| Effective electric charge of bunch head | 196pC |
| Bunch length(FWHM) | 1.3ps |
| Bunch radius | 2.8mm |
| Energy spread(FWHM) | 0.28% |
| Energy spread (rms) of effective bunch head | 0.27% |
| Transverse normalized emittance (rms) | 6.5π mm·mrad |

## 4  Beam tail effect analysis

### 4.1  Beam dynamics of multi-pulses

As one of merits of EC-ITC RF gun mentioned in Ref. [9,10], the BB power can be eliminated to almost zero, which was confirmed by Parmela as well. However, for providing stable high-performance bunches required by FEL injector, it's critical to analyze the effects of beam tail behind effective bunch head and interactions of multi-pulses. During the whole simulation, the calculation of space charge field is of great importance to space-charge dominated beam with high intensity and low energy. A cylindrical grid consisting of rings in the radial direction and slices in the longitudinal direction is set up over the extension of the bunch, particle coordinates and momentums are Lorentz-transformed into the average rest frame of the space-charge mesh, and a static field calculation can be performed by integrating numerically over the rings thereby assuming a constant charge density inside a ring. Then the field contributions of the individual rings at the center points of the grid cells are added up and transformed back into the lab frame[15,16]. For demonstration, a double-pulse case is taken as example and shown in Fig. 6.

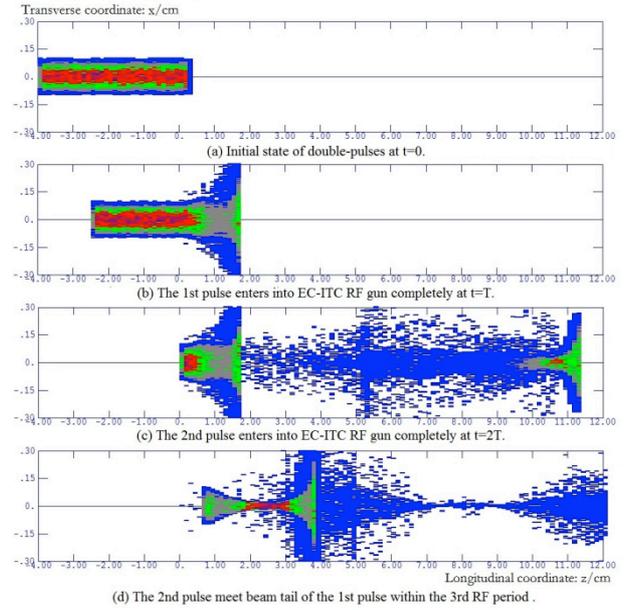

Fig. 6. Beam envelopes of double-pulses in EC-ITC-RF gun at different times.

From Fig. 6(c) and (d), obviously, the influences of the first pulse's tail couldn't be ignored at all. For simplification, the output of first pulse can be conveniently described as three parts: electrons of Part-1 are bunched and compose the effective bunch head, and electrons of Part-2 have deviated energy from bunch head and will be extracted together with Part-1, while Part-3, so called beam tail, encounter the re-verse of accelerating electric field and don't have enough kinetic energy to conquer it in Cell-2, thus they are decelerated, even retrograde and bombard the second bunch head. Both Part-2 and Part-3 are unavoidable, then it's necessary to analyze the uncertain influences of them on effective bunch heads, which might encumber the properties of FEL injector.

### 4.2 Interactions of multi-pulses

Considering only one reference particle used in Parmela simulation, the explicit properties of multi-bunches couldn't be calculated and given directly. It's so necessary to solve this issue, that a post-processing MATLAB code has been written based on beam dynamic results and following statistical formulas of rms emittance and rms energy spread.

$$\varepsilon_{x,\text{rms}} = \frac{\beta\gamma}{N}\sqrt{\sum_{i=1}^{N} x_i^2 \sum_{i=1}^{N} x_i'^2 - \left(\sum_{i=1}^{N} x_i x_i'\right)^2}. \quad (4)$$

Where, $\beta$ and $\gamma$ are relative velocity and relative kinetic energy of the effective bunch, while $x_i$ and $x_i'$ are the transverse coordinate and divergence coordinate of the $i$th particle of output effective bunch separately.

$$\Delta W_{\text{rms}} = \frac{1}{\overline{W}}\sqrt{\frac{1}{N}\sum_{i=1}^{N}\left(W_i - \overline{W}\right)^2}. \quad (5)$$

Where, $W_i$ is the absolute energy of the $i$th particle and $\overline{W}$ is the mean energy of effective bunch.

For the sake of accuracy and reliability, the case of four pulses is simulated. The phase / energy spectra as well as transverse phase spaces are shown in Fig. 7.

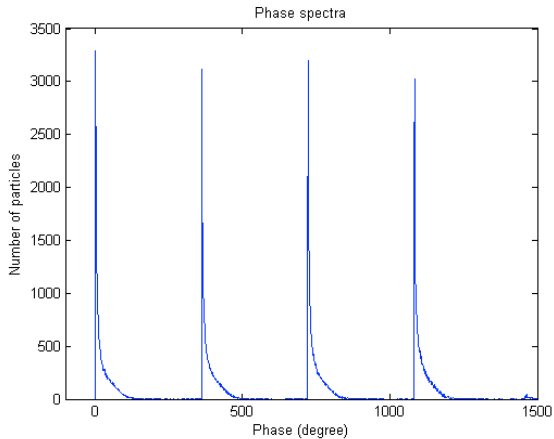

(a) Phase spectra of 4-bunches.

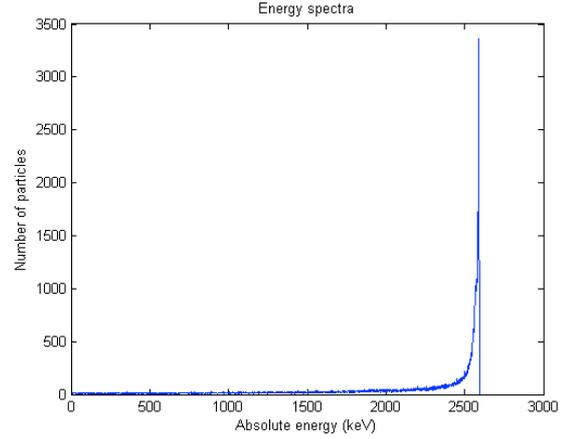

(b) Energy spectra of 4-bunches.

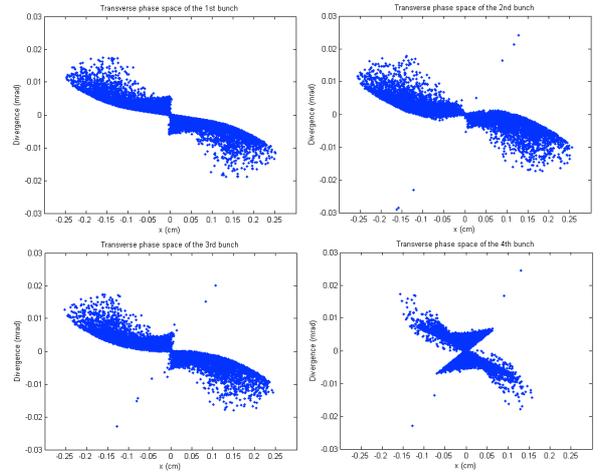

(c) Transverse phase spaces of 4-bunches.
Fig. 7. Bunch properties with 4-pulses input.

For instance, specific parameters of the second bunch are calculated and listed in Table 3.

Table 3. Specific parameters of the 2nd bunch.

| Parameter | Value |
| --- | --- |
| Beam kinetic energy | 2.59MeV |
| Effective electric charge of bunch head | 242pC |
| Bunch length(FWHM) | 2.2ps |
| Bunch radius | 2.8mm |
| Energy spread(FWHM) | 0.27% |
| Energy spread (rms) of effective bunch head | 0.32% |
| Transverse normalized emittance (rms) | 5.1$\pi$mm·mrad |

Though the rms energy spread is deteriorated slightly, the results demonstrate a significant benefit that the effective electric charge of bunch head has a 20% increase, to 242pC. This effect is important for performance enhance of the FEL injector.

In addition, main properties of each bunch are listed in Table 4 separately, which indicate that stable high brilliant electron bunch source can be obtained.

Table 4. Main properties of each bunch with 4-pulses input.

| Parameter | The 1st bunch | The 2nd bunch | The 3rd bunch | The 4th bunch |
| --- | --- | --- | --- | --- |
| Effective electric charge of bunch head | 196pC | 242pC | 233pC | 237pC |
| Energy spread (rms) of effective bunch head | 0.27% | 0.32% | 0.28% | 0.32% |
| Transverse normalized emittance (rms) | 6.5$\pi$mm·mrad | 5.1$\pi$mm·mrad | 4.1$\pi$mm mrad | 3.8$\pi$mm·mrad |
| Bunch length(FWHM) | 1.3ps | 2.2ps | 2.2ps | 3.1ps |

### 4.3 Beam tail effects

To interpret the phenomena summarized as above, particle tracking and analysis codes are used here, which give all output particles distributions with independent phases and energies. Parts of low energy particles in beam tail fall behind and coincide with the second bunch head on both phase and energy, then they are accelerated and exported together. Nevertheless, the energy deviations between low energy tail of the first beam and bunch head of the second beam still exist and deteriorate energy spread slightly. However, the second bunch head still has brilliant performances matching with FEL injector demands.

For comparison, both phase spectra with and without space charge force are illustrated by Fig. 8 separately. In the case of space charge effects included, though main bunch strength is lower, sub-bunch charge is larger due to lengthening effect of space charge field.

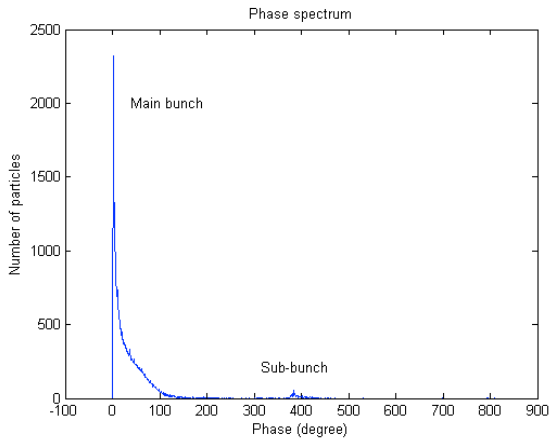

(a) Phase spectrum with space charge force.

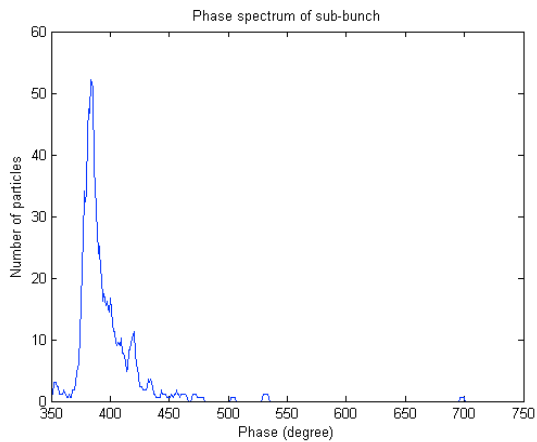

(b) Phase spectrum of sub-bunch with space charge force.

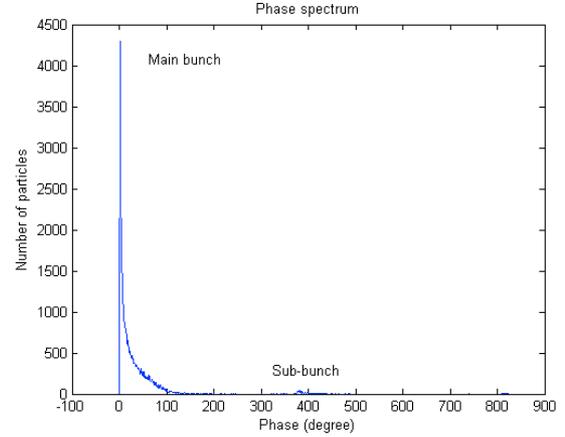

(c) Phase spectrum without space charge force.

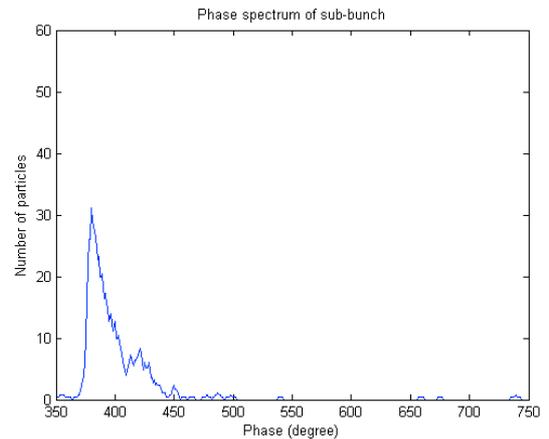

(d) Phase spectrum of sub-bunch without space charge force.

Fig. 8. Phase spectrum of the whole extracted beam containing main-bunch and sub-bunch with single pulse input.

Observed from Fig. 8, besides low energy tail, unexpected energy-deviated particles are mixed and extracted together with effective bunch head, which couldn't be eliminated in ITC-RF gun. After transmitting into the following traveling-wave accelerator with focusing coils, energy-deviated particles would be separated from effective bunch head during accelerating process. Output beam phase spectrum is plotted in Fig. 9, which demonstrates that energy-deviated particles are separated to two small bunches behind effective bunch head, and extracted with the second and third bunch head in the following two RF periods, respectively. Additionally, note that only 6.52% particles mix with the second bunch head, and extracted energy of them are about 2MeV less than effective bunch head, which means the energy-deviated particles might be lost in transmission system with bending magnets.

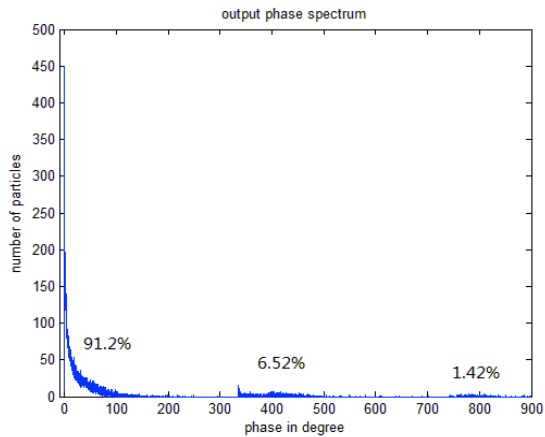

Fig. 9. Phase spectrum of extracted beam from traveling-wave accelerator.

## 5 Conclusions

Aiming at taking advantages of simplification and compactness of thermionic RF gun rather than complicated and expensive photocathode RF gun, a type of performance-enhanced EC-ITC RF gun is proposed. multi-cavity structure theories and EC-ITC RF gun criterions are combined therein. And the beam dynamics indicate that it can generated high quality bunches with ~200pC effective bunch charge, which is very suitable for THz-FEL.

Nevertheless, in order to guarantee the continual stability of extracted effective bunches with high quality, beam tail effects and interactions mentioned above can't be neglected which are resulted from the non-independence of multi-bunches. The analysis results using homemade MATLAB codes reveal that the beam tail consists of low energy electrons would contribute to the next bunch head, and make the effective bunch charge increase to over 15% more than before while the energy spread only deteriorates slightly. Besides, joint debugging with traveling-wave accelerator illustrates that energy-deviated electrons would be separated from bunch head and compose the sub-bunch in the next RF period with only 6.52% particles, that could be lost in the following transmission system since their energy are about 2MeV less than bunch head.

In summary, beam tail effects are observed and described in details by means of numerical analysis. And the simulation results indicate that continual stable bunches could be obtained, which have improved our confidence to a great extent in adopting performance-enhanced EC-ITC RF gun for THz-FEL injectors.